\documentclass[twocolumn,showpacs,aps,pre]{revtex4}

\usepackage{amssymb,amsmath}
\usepackage[dvips]{graphicx}

\begin{document}

\title{Phase transitions of the generalized contact process with two absorbing states}

\author{Man Young Lee}
\author{Thomas Vojta}
\affiliation{Department of Physics, Missouri University of Science and Technology, Rolla, MO 65409, USA}

\begin{abstract}
We investigate the generalized contact process with two absorbing states in one space
dimension by means of large-scale Monte-Carlo simulations. Treating the creation rate of
active sites between inactive domains as an independent parameter leads to a rich phase
diagram. In addition to the conventional active and inactive phases we find a parameter
region where the simple contact process is inactive, but an \emph{infinitesimal} creation
rate at the boundary between inactive domains is sufficient to take the system into the
active phase. Thus, the generalized contact process has two different phase transition
lines. The point separating them shares some characteristics with a multicritical point.
We also study in detail the critical behaviors of these transitions and their
universality.
\end{abstract}

\date{\today}
\pacs{05.70.Ln, 64.60.Ht, 02.50.Ey}

\maketitle

\section{Introduction}

Many systems in physics, chemistry, and biology are far from thermal equilibrium,
even if they are in time-independent steady states. In recent years, continuous
phase transitions between different nonequilibrium steady states have attracted
lots of attention. Just as in equilibrium, these transitions are characterized
by large-scale fluctuations and collective behavior over large distances and long
times.
Examples can be found, e.g., in surface growth, granular flow, chemical reactions,
population dynamics, and even in traffic jams
\cite{ZhdanovKasemo94,SchmittmannZia95,MarroDickman99,Hinrichsen00,Odor04,Luebeck04,TauberHowardVollmayrLee05}.

Continuous nonequilibrium phase transitions can be divided into different universality
classes according to their critical behavior, and considerable effort has been devoted
to categorizing the variety of known transitions. A well-studied type of
nonequilibrium phase transitions separates fluctuating (active) steady states from absorbing
(inactive) states where fluctuations stop completely. The generic universality class for
these so-called absorbing state transitions is directed percolation (DP) \cite{GrassbergerdelaTorre79}.
More specifically, it was conjectured by Janssen and Grassberger \cite{Janssen81,Grassberger82}
that all absorbing state transitions with a scalar order parameter and short-range interactions
belong to this class as long as there are no extra symmetries or conservation laws.
While nonequilibrium transitions in the DP universality class are ubiquitous in both theory and
computer simulations, experimental verifications were only found rather recently
in ferrofluidic spikes \cite{RuppRichterRehberg03} and in the transition between two
turbulent states in a liquid crystal \cite{TKCS07}.

Absorbing state transitions in universality classes different from DP can occur in the
presence of additional symmetries or conservation laws. Hinrichsen \cite{Hinrichsen97}
introduced nonequilibrium lattice models with $n\ge 2$ absorbing states. In the case of
two symmetric absorbing states ($n=2$), he found the transition to be in a new
universality class, the $Z_2$-symmetric directed percolation class (DP2). If the symmetry
between the absorbing states is broken, the critical behavior reverts back to DP. In one
dimension, the DP2 universality class coincides \cite{Hinrichsen00} with the
parity-conserving PC class \cite{GrassbergerKrauseTwer84} which is observed, e.g., in the
branching-annihilating random walk with an even number of offspring (BARWE)
\cite{ZhongAvraham95}.

In this paper, we revisit one of the stochastic lattice models introduced in Ref.\
\cite{Hinrichsen97}, the generalized contact process with two absorbing states in one
space dimension.
Compared to the simple contact process \cite{HarrisTE74}, this model contains an
additional dynamical process, \emph{viz.}, the creation of active sites at the boundary
between domains of different inactive states. By treating the rate for this process as an
independent parameter we uncover a rich phase diagram with two different types
of phase transitions, separated by a special point that shares many characteristics
with a multicritical point. We perform large-scale Monte-Carlo
simulations of this model to study in detail the critical behavior of these transitions.

Our paper is organized as follows. We introduce the generalized contact process with
several absorbing states in Sec.\ \ref{sec:processes}. In Sec.\  \ref{sec:mf}, we
summarize the mean-field theory for this system. Sec.\ \ref{sec:simulations} is devoted
to the results and interpretation of our Monte-Carlo simulations. We conclude in Sec.\
\ref{sec:conclusions}.

\section{The generalized contact process with several absorbing states}
\label{sec:processes}

The contact process \cite{HarrisTE74} is a paradigmatic model in the DP universality class. It is
defined on a $d$-dimensional hypercubic lattice. Each lattice site $\mathbf{r}$ can be in
one of two states, namely A, the active (infected) state or I, the inactive (healthy)
state. Over the course of the time
evolution, active sites can infect their nearest neighbors, or they can become inactive
spontaneously. More precisely, the contact process is a continuous-time Markov process
during which active sites turn inactive at a rate $\mu$, while inactive sites become
infected at a rate $\lambda m/(2d)$ where $m$ is the number of active nearest neighbors.
The healing rate $\mu$ and the infection rate $\lambda$ are external parameters whose
ratio determines the behavior of the system.

If $\mu \gg \lambda$, healing dominates over infection. All infected sites
will eventually become inactive, leaving the absorbing state without any active sites the
only steady state. Thus, the system is in the inactive phase. In the opposite limit,
$\lambda \gg \mu$, the infection survives for infinite times, i.e.,
there is a steady state with a nonzero density of active sites. This is the active phase.
The nonequilibrium phase transition between these two phases at a critical value of the ratio
$\lambda/\mu$ is in the DP universality class.

In 1997, Hinrichsen \cite{Hinrichsen97} introduced a generalization of the contact
process. Each lattice site can now be in one of $n+1$ states, the active
state A or one of  the $n$ different inactive states I$_k$ ($k=1\ldots n$).
$k$ is sometimes called the ``color'' index. The dynamics
of the generalized contact process is defined via the following rates for transitions of
pairs of nearest-neighbor sites,
\begin{eqnarray}
w(\textrm{AA} \to \textrm{AI}_k) = w(\textrm{AA} \to \textrm{I}_k\textrm{A}) &=& \bar\mu/n~,
\label{eq:rate_barmu}\\
w(\textrm{AI}_k \to \textrm{I}_k\textrm{I}_k) = w(\textrm{I}_k\textrm{A} \to \textrm{I}_k\textrm{I}_k) &=& \mu_k~,\\
w(\textrm{AI}_k \to \textrm{AA}) = w(\textrm{I}_k\textrm{A} \to \textrm{AA}) &=& \lambda~,
\label{eq:rate_lambda}\\
w(\textrm{I}_k\textrm{I}_l \to \textrm{I}_k\textrm{A}) = w(\textrm{I}_k\textrm{I}_l \to \textrm{A}\textrm{I}_l) &=&
\sigma~,
\label{eq:rate_sigma}
\end{eqnarray}
with $k,l=1\ldots n$ and $k \ne l$. All other rates vanish. We are mostly interested in
the fully symmetric case, $\mu_k \equiv \mu$ for all $k$. For $n=1$ and $\bar \mu = \mu$,
the so defined generalized contact process coincides with the simple contact process
discussed above. One of the rates $\bar \mu, \mu, \lambda$, and $\sigma$ can be set to
unity without loss of generality, thereby fixing the unit of time. We choose $\lambda=1$
in the following. Moreover, to keep the parameter space manageable, we focus on the case
$\bar \mu =\mu $ in the bulk of the paper. The changes for $\bar \mu \ne \mu$ will be
briefly discussed in Sec.\ \ref{sec:conclusions}.

The process (\ref{eq:rate_sigma}) prevents inactive domains of different color
(different $k$) to stick together indefinitely. They can separate, leaving active sites
in between. Thus, this transition allows the domain walls to move through space.
It is important to realize that without the process (\ref{eq:rate_sigma}), i.e., for $\sigma=0$,
the color of the inactive sites becomes unimportant, and all $\textrm{I}_k$ can be
identified. Consequently, for $\sigma=0$, the dynamics of the generalized contact process reduces
to that of the simple contact process for all values of $n$.

Hinrichsen \cite{Hinrichsen97} studied the one-dimensional generalized contact process
by means of Monte-Carlo simulations, focusing on the case $\sigma=\lambda=1$.
For $n=2$, he found a nonequilibrium phase transition at a finite value of $\mu$ which
separates the active and inactive phases. The critical behavior of this transition
coincides with that of the PC universality class. For $n \ge 3$, he found the model to be
always in the active phase. The Monte-Carlo simulations were later confirmed by means of
a non-hermitian density-matrix renormalization group study \cite{HooyberghsCarlonVanderzande01}.

Motivated by a seeming discrepancy between these results and simulations that we
performed during our study of absorbing state transitions on a percolating lattice
\cite{LeeVojta09}, we revisit the one-dimensional generalized contact process
with two inactive states. In contrast to the earlier works we treat the rate $\sigma$ of the process
(\ref{eq:rate_sigma}) as an independent parameter (rather than fixing it at
$\sigma=\lambda=1$).

\section{Mean-field theory}
\label{sec:mf}

To get a rough overview over the behavior of the generalized contact process with
two inactive states, we first
perform a mean-field analysis.  Denoting the probabilities for a site to be in state
A, I$_1$, and I$_2$ with $P_A$, $P_1$, and $P_2$, respectively, the mean-field equations
read:
\begin{eqnarray}
dP_A / dt &=& (1-\mu) P_A  - P_A^2 + 2 \sigma P_1 P_2~,
\label{eq:mf_PA}\\
dP_1 / dt &=& \mu P_A /2 - P_A P_1 - \sigma P_1 P_2~,
\label{eq:mf_P1}\\
dP_2 / dt &=& \mu P_A /2 - P_A P_2 - \sigma P_1 P_2~.
\label{eq:mf_P2}
\end{eqnarray}
Let us begin by discussing the steady states which are given by the
fixed points of the mean-field equations. There are two trivial, inactive fixed points
$P_1=1, P_A=P_2=0$ and $P_2=1, P_A=P_1=0$. They exist for all values of the parameters
$\mu$ and $\sigma$ and correspond to the two absorbing states. In the case of $\sigma=0$, these
fixed points are unstable for $\mu<1$ and stable for $\mu>1$. In contrast, for $\sigma
>0$, they are always unstable.

The active fixed point is given by $P_1=P_2$ and fulfills the equation
\begin{equation}
0= (1-\mu) P_A - P_A^2 + \sigma (1-P_A)^2 /2~.
\label{eq:mf_PA_steady}
\end{equation}
For $\sigma=0$, this equation reduces to the well-known mean-field equation of the simple
contact process, $0=(1-\mu) P_A -P_A^2$ with the solution $P_A=1-\mu$ for $\mu<1$. Thus,
for $\sigma=0$, the nonequilibrium phase transition of the generalized contact process
occurs at $\mu=\mu_c^{cp}=1$. This means, it coincides with the transition of the simple
contact process, in agreement with the general arguments given in Sec.\
\ref{sec:processes}. In the general case, $\sigma \ne 0$, the steady state density of
active sites, $P_A$, is given by the positive solution of
\begin{equation}
P_A = \frac 1 {2-\sigma} \left(1-\mu-\sigma \pm \sqrt{\mu^2 -2\mu +1 +2 \mu \sigma}
\right)~.
\label{eq:mf_solution_PA}
\end{equation}
We are particularly interested in the behavior of $P_A$ for small $\sigma$. As long as
$\mu<\mu_c^{cp}=1$ (i.e., in the active phase of the simple contact process), a small, nonzero
$\sigma$ only provides a subleading correction to $P_A$. At $\mu=\mu_c^{cp}=1$,
the density of active sites vanishes as $P_A \sim \sqrt{\sigma}$ with $\sigma \to 0$.
Finally, for $\mu>\mu_c^{cp}=1$, the density of active sites vanishes as $P_A \sim
\sigma/(\mu-1)$.

We thus conclude that within mean-field theory, the generalized contact process with two
inactive states is in the active phase for any nonzero $\sigma$. This agrees with older
mean-field results but disagrees with more sophisticated methods
which predict a nonequilibrium transition at a finite value of $\mu$ \cite{Hinrichsen97,HooyberghsCarlonVanderzande01}.
The mean-field dynamics can be worked out in a similar fashion. We find that the approach
to the stationary state is exponential in time anywhere in parameter space except for the
critical point of the simple contact process at $\mu=1, \sigma=0$. However, it is known
that mean-field theory does not reflect the correct long-time dynamics of the generalized
contact process which is of power-law type \cite{Hinrichsen97}. Therefore, we do not
analyze the mean-field dynamics in detail.

\section{Monte Carlo simulations}
\label{sec:simulations}

\subsection{Method and overview}
\label{subsec:method}

We now turn to the main part of the paper, \emph{viz.}, large-scale Monte-Carlo
simulations of the one-dimensional generalized contact process with two inactive states.
We perform two different types of calculations: (i) decay runs and (ii) spreading runs.
Decay runs start from a completely active lattice; we monitor the time evolution of the
density $\rho(t)$ of active sites as well as the densities $\rho_1(t)$ and $\rho_2(t)$ of
sites in inactive states I$_1$ and I$_2$, respectively. Spreading simulations start from
a single active (seed) site embedded in a system of sites in state I$_1$. (From a domain
wall point of view, the spreading runs are therefore in the even parity sector.) Here we
measure the survival probability $P_s(t)$, the number of sites in the active cloud
$N_s(t)$ and the mean-square radius of this cloud, $R^2(t)$.

In each case, the simulation proceeds as a sequence of events. In each event,
a pair of nearest-neighbor sites is
randomly selected from the active region. For the spreading simulations, the active
region initially consists of the seed site and its neighbors; it is updated in the course
of the simulation according to the actual size of the active cluster. For the decay runs,
the active region comprises the entire sample. The selected pair than undergoes one of the possible
transitions according to eqs.\ (\ref{eq:rate_barmu}) to (\ref{eq:rate_sigma}) with
probability $\tau w$. Here the time step $\tau$ is a constant which we have fixed at 1/2.
The time increment associated with the event is $\tau/N_{pair}$ where $N_{pair}$
is the number of nearest-neighbor pairs in the active region.

\begin{figure}[tb]
\includegraphics[width=\columnwidth]{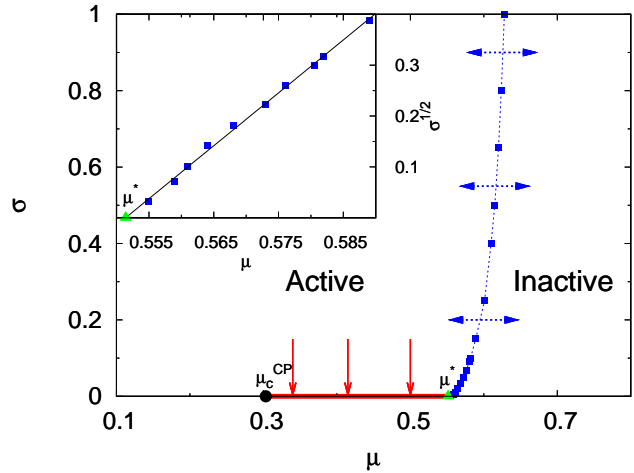}
\caption{(Color online) Phase diagram of the 1D generalized contact process
as function of the healing rate $\mu$ and the boundary rate $\sigma$. A line of
DP2 (PC) transitions (blue dashed line) separates
the active and inactive phases. For $\sigma \to 0$, this line does not terminate in
the simple contact process critical point at $\mu_c^{cp}\approx 0.30325$ and but
at  $\mu^* \approx 0.552$.  For $\mu_c^{cp} < \mu < \mu^*$,
the system is inactive at $\sigma=0$ (thick solid red line), but an infinitesimal
$\sigma$ takes it to the active phase.
Inset: Close to the endpoint at $\mu^*$, the phase boundary behaves roughly as
$\sigma_c \sim (\mu-\mu^*)^2$.}
\label{fig:pd}
\end{figure}

Using this method we studied systems with sizes up to $L=10^6$ lattice sites and times
up to $t_{max}=10^8$, exploring the parameter space $0\le\mu\le1$ and $0\le\sigma\le1$.
The $\sigma-\mu$ phase diagram resulting from our simulations is displayed in Fig.\
\ref{fig:pd}.
This phase diagram shows that the crossover from DP critical behavior at $\sigma=0$ to
DP2 (or, equivalently, PC) critical behavior at $\sigma >0$ occurs in an unusual fashion.
The phase boundary $\sigma_c(\mu)$ between the active and inactive phases does not
terminate at the critical point of the simple contact process located at
$(\mu,\sigma)=(\mu_c^{cp},0)\approx (0.30325,0)$. Instead, it ends at the point
$(\mu,\sigma)=(\mu^*,0)\approx (0.552,0)$. In the parameter range $\mu_c^{cp} < \mu <
\mu^*$, the system is inactive at $\sigma=0$, but an infinitesimally small nonzero
$\sigma$ takes it to the active phase.

Thus, the one-dimensional generalized contact process with two inactive states
has two types of phase transitions, (i) the generic transition occurring at $\mu>\mu^*$
and $\sigma=\sigma_c(\mu)>0$ (marked by the dashed blue line and arrows in Fig. \ref{fig:pd})
and (ii) the transition occurring for $\mu_c^{cp}<\mu<\mu^*$ as $\sigma$ approaches
zero (solid red line and arrows). We note in passing that our critical
healing rate for $\sigma=1$ is $\mu_c=0.628(1)$, in agreement with Ref.\
\cite{Hinrichsen97}

In the following subsections we first discuss in detail the simulations that lead to this
phase diagram, and then we present results on the critical behavior of both transitions
as well as special point $(\mu^*,0)\approx (0.552,0)$ that separates them.

\subsection{Establishing the phase diagram}
\label{subsec:results}

\begin{figure}[tb]
\includegraphics[width=\columnwidth]{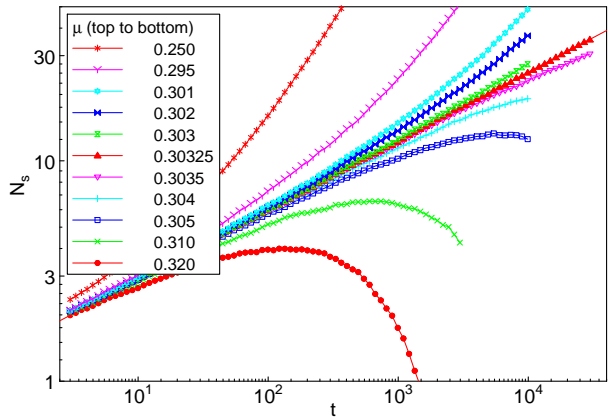}
\caption{(Color online) Spreading simulations at $\sigma=0$: Number $N_s$ of active sites
as a function of time $t$. The solid line for $\mu=0.30325$ represents a fit to
$N_s \sim t^{\Theta_{cp}}$ yielding $\Theta_{cp}=0.315(5)$. The data are averages over 25000 runs.}
\label{fig:na_clean}
\end{figure}

We first performed a number of spreading simulations at $\sigma=0$ and various $\mu$
for maximum times up to $3\times 10^4$. The resulting number
$N_s(t)$ of active sites in the cluster is shown in Fig.\ \ref{fig:na_clean}.
The figure demonstrates that the transition between the active and inactive phases occurs at
$\mu = 0.30325(25)$. A fit of the critical curve to $N_s \sim t^{\Theta_{cp}}$ yields
$\Theta_{cp}=0.315(5)$. As expected from the general arguments in Sec.\ \ref{sec:processes},
both the critical healing rate and the initial slip exponent $\Theta_{cp}$ agree very well with
the results of the simple contact process (see, e.g., Ref.\ \cite{Jensen99} for accurate
estimates of the DP exponents). Thus, at $\sigma=0$, the generalized
contact process undergoes a transition in the directed percolation universality class
at $\mu = \mu_c^{cp}=0.30325(25)$.

\begin{figure}
\includegraphics[width=\columnwidth]{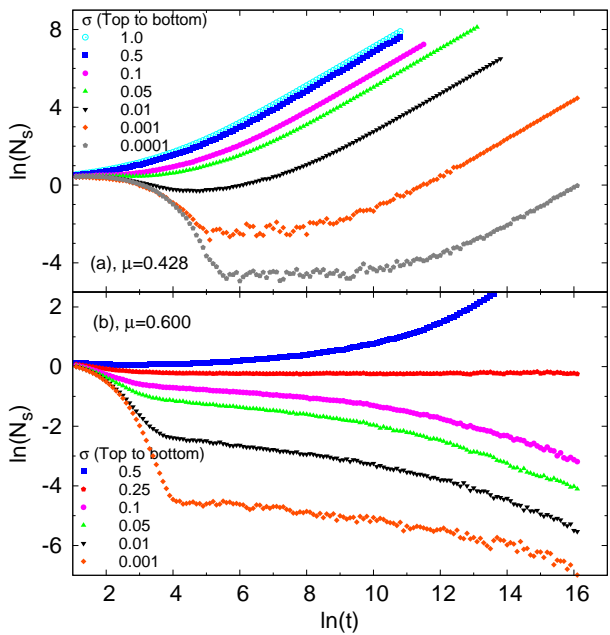}
\caption{(Color online) Spreading simulations: Number $N_s$ of active sites as
a function of time $t$  for several $\sigma$ at fixed $\mu=0.428$ (panel a) and
$\mu=0.6$ (panel b). The data are averages over $10^3$ (at the smallest $\sigma$) to $10^5$ runs.}
\label{fig:Ns(t)_sigma}
\end{figure}

We now turn to nonzero $\sigma$. Because the domain boundary process
(\ref{eq:rate_sigma}) creates extra active sites, it is clear that the phase boundary
between the active and inactive phases has to shift to larger healing rates $\mu$
with increasing $\sigma$. In the simplest crossover scenario, the phase boundary
$\sigma_c(\mu)$ would behave as $\sigma_c \sim (\mu-\mu_c^{cp})^{1/\phi}$ where $\phi$ is a
crossover exponent. To test this scenario, we performed spreading simulations for times up
to $10^7$ at several
fixed $\mu > \mu_c^{cp}$ in which we vary $\sigma$ to locate the transition.
Examples of the
resulting $N_s(t)$ curves for several $\sigma$ at $\mu=0.428$ and $\mu=0.6$
are shown in Fig.\ \ref{fig:Ns(t)_sigma}.
The set of curves for $\mu=0.6$ (Fig.\ \ref{fig:Ns(t)_sigma}b) behaves as expected: Initially, $N_s(t)$
follows the behavior of the simple contact process at this $\mu$. At later
times, the curves with $\sigma \gtrapprox 0.25$ curve upwards implying that
the system is in the active phase.
The curves for $\sigma \lessapprox 0.25$ curve downward, indicating that the system
is in the inactive phase. Thus, $\sigma_c(\mu=0.6) \approx 0.25$.

In contrast, the set of curves for $\mu=0.428$ (Fig.\ \ref{fig:Ns(t)_sigma}a) behaves
very differently. After an initial decay, $N_s(t)$ curves strongly upwards for all values
of $\sigma$ down to the smallest value studied, $\sigma=10^{-4}$. This suggests that
at $\mu=0.428$, any nonzero $\sigma$ takes the generalized contact process to the active
phase. The phase transition thus occurs at $\sigma=0$.

We determined analogous sets of curves for many different values of the healing rate in the
interval $\mu_c^{cp}=0.30325 < \mu < 0.65$. We found that the phase transition to the
active phase occurs at $\sigma=0$ for $\mu_c^{cp} < \mu < \mu^* =0.552$, while it occurs
at a nonzero $\sigma$ for healing rates $\mu > \mu^*$. This establishes the phase diagram
shown in Fig.\ \ref{fig:pd}. The phase boundary thus does \emph{not} follow the simple crossover
scenario outlined above. In the following subsections, we analyze in detail the critical
behavior of the different nonequilibrium phase transitions.

\subsection{Generic transition}
\label{subsec:generic}

\begin{figure}
\includegraphics[width=\columnwidth]{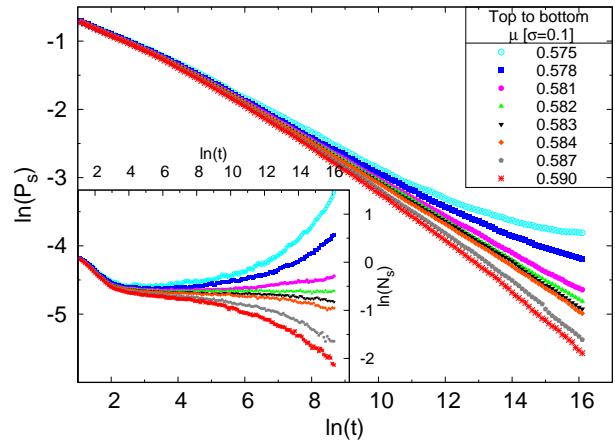}
\caption{(Color online) Spreading simulations at $\sigma=0.1$ for several $\mu$ close to the phase boundary.
Main panel: Number $N_s$ of active sites as a function of time $t$. Inset: Survival probability $P_s$
as a function of time $t$. The data are averages over $10^5$ runs.}
\label{fig:DP2_transition}
\end{figure}

\begin{figure}
\includegraphics[width=\columnwidth]{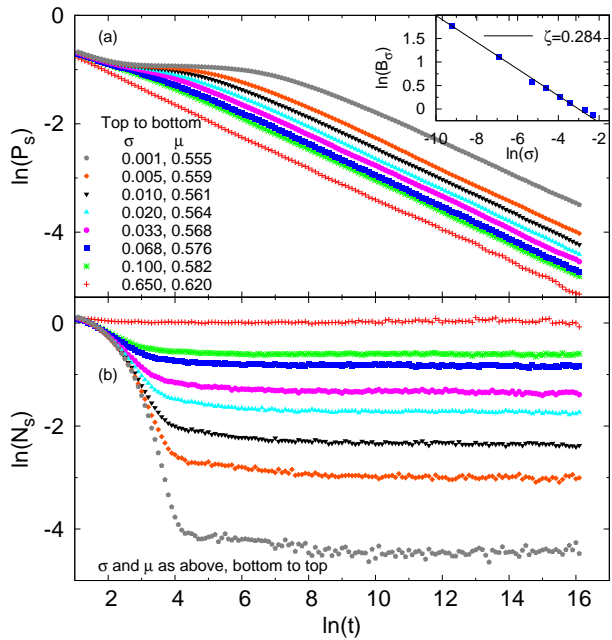}
\caption{(Color online) Critical spreading simulations:  Survival probability $P_s$ and number of
          active sites $N_s$ as functions of $t$ for several points $(\mu,\sigma)$ located on the
          generic phase boundary. The inset shows the prefactor $B_\sigma$ of the critical power
          law $P_s = B_\sigma t^{-\delta}$ as a function of $\sigma$. The solid line is a fit to
          $B_\sigma \sim \sigma^{-\zeta}$ which gives $\zeta=0.284$.}
\label{fig:all_critical}
\end{figure}

We first consider the generic transition occurring at $\mu > \mu^* \approx 0.552$ and nonzero
$\sigma$ (the blue dashed line in Fig.\ \ref{fig:pd}). Figure \ref{fig:DP2_transition}
shows a set of spreading simulations at $\sigma=0.1$ and several $\mu$ in the vicinity of the phase boundary.
The data indicate a critical point at $\mu\approx 0.582$. We performed analogous
simulations for several points on the phase boundary. Figure \ref{fig:all_critical}
shows the survival probability $P_s$ and number $N_s$ of active sites as functions of time for all
the respective critical points.
In log-log representation, the $N_s$ and $P_s$ curves for different $\sigma$ and $\mu$
are perfectly parallel, i.e., they represent power-laws  with the same exponent.
Fits of the asymptotic long-time behavior to $P_s =B_\sigma  t^{-\delta}$ and
$N_s = C_\sigma t^{\Theta}$ give estimates of
$\delta=0.289(5)$ and $\Theta=0.000(5)$. Moreover, we measured (not shown)
the mean-square radius $R^2(t)$ of the active cloud
as a function of time. Its long time behavior follows a universal power law.
Fitting to $R^2(t) \sim t^{2/z}$ gives $2/z=1.145(5)$ ($z=1.747(7)$).
Here $z=\nu^\parallel /\nu^\perp$ is the dynamical exponent, i.e., the ratio
between the correlation time exponent $\nu_\parallel$ and the correlation length
exponent $\nu_\perp$.

\begin{figure}
\includegraphics[width=\columnwidth]{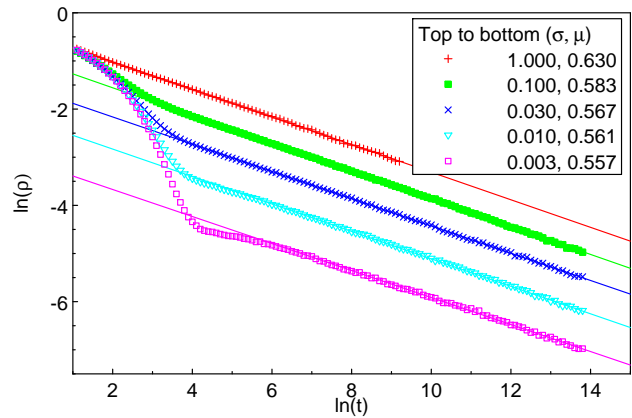}
\caption{(Color online) Critical density decay simulations:  Density $\rho_A$ of active sites
          as function of $t$ for several points $(\mu,\sigma)$ on the
          generic phase boundary. The solid lines are fits to a power law $\rho_A = \bar B_\sigma t^{-\alpha}$
          giving $\alpha=0.285(5)$. The data represent averages of 400 runs with system size
          $L=10^4$.}
\label{fig:density_critical}
\end{figure}

In addition to the spreading simulations, we also performed density decay simulations for
several $(\mu,\sigma)$ points on the phase boundary. Characteristic results are presented in
Fig.\ \ref{fig:density_critical}.
The figure shows that the density $\rho_A$ of active sites at criticality follows a universal
power law, $\rho_A = \bar B_\sigma t^{-\alpha}$ at long times. The corresponding fits give $\alpha=0.285(5)$
which agrees (within the error bars) with our value of the survival probability exponent $\delta$.
We thus conclude that the generic transition of our system is characterized by three independent
exponents (for instance $\nu_\perp, z$ and  $\delta$) rather than four (as could be expected for a
general absorbing state transition \cite{Hinrichsen00}).
We point out, however, that even though $P_s$ and $\rho_A$ show the same power-law time
dependence at criticality, the behavior of the prefactors differs. Specifically, the prefactor $\bar B_\sigma$
of the density is increasing with increasing $\sigma$ while the prefactor $B_\sigma$ of the survival probability
decreases with increasing $\sigma$.

All the exponents of the generic transition do not depend on $\mu$ or $\sigma$, implying that the critical behavior
is universal.  Moreover, their values are in excellent agreement with the known values of the PC (or DP2)
universality class (see, e.g., Ref.\ \cite{Hinrichsen00,Odor04}). We therefore conclude that the critical behavior of the
generic transition of generalized contact process with two inactive states is universally in this class.

\subsection{Transition at $\sigma=0$}
\label{subsec:sigma0}

After discussing the generic transition, we now turn to the line of transitions
at $\mu_c^{cp} < \mu <\mu^*$ and $\sigma=0$. To investigate these transitions more
closely, we performed both spreading and density decay simulations at fixed $\mu$
and several $\sigma$-values approaching $\sigma=0$ (as indicated by the solid (red) arrows
in the phase diagram, Fig.\ \ref{fig:pd}).

\begin{figure}
\includegraphics[width=\columnwidth]{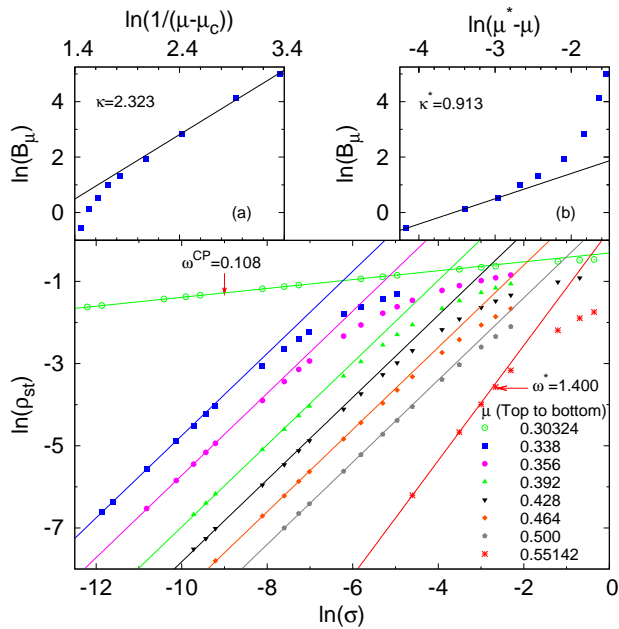}
\caption{(Color online) Density decay simulations. Main panel: stationary density $\rho_{st}$ as a function
of the boundary rate $\sigma$ for various healing rates $\mu$. For $\mu_c^{cp} < \mu <\mu^*$,
the solid lines are fits of the low-$\sigma$ behavior to
$\rho_{st} = B_\mu \sigma$. At the simple contact process critical point, $\mu=\mu_c^{cp}=0.30324$,
and at the endpoint, $\mu=\mu^*=0.552$, we fit to power-laws
$\rho_{st} \sim \sigma^{\omega}$ which gives exponents of $\omega_{cp} = 0.108(2)$ and $\omega^*=1.4(1)$.
The data are averages over 50 to 200 runs with system sizes $L=2000$ to 5000.
Inset a: prefactor $B_\mu$ of the linear $\sigma$
dependence
as a function of $\mu-\mu_c^{cp}$. A fit to a power law gives $B_\mu \sim (\mu-\mu_c^{cp})^{-\kappa}$
with $\kappa=2.32(10)$. Inset b: prefactor $B_\mu$
as a function of $\mu^*-\mu$. A fit to a power law gives $B_\mu \sim (\mu^*-\mu)^{\kappa^*}$
with $\kappa^*=0.91$.}
\label{fig:rho_st}
\end{figure}

Let us start by discussing the density decay simulations.
Figure \ref{fig:rho_st} shows the stationary density
$\rho_{st}$ of active sites as a function of $\sigma$ for several values of the healing rate
$\mu$.
Interestingly, the stationary density depends linearly on $\sigma$ for all healing rates
$\mu_c^{cp} < \mu <\mu^*$, in seeming agreement with mean-field theory.
This means $\rho_{st} = B_\mu
\sigma^\omega$ with $\omega=1$ and $B_\mu$ being a $\mu$-dependent constant.
We also analyzed, how the prefactor $B_\mu$ of the mean-field-like  behavior depends on
the distance from the simple contact process critical point. As inset a) of
Fig.\ \ref{fig:rho_st} shows, $B_\mu$ diverges as $(\mu-\mu_c^{cp})^{-\kappa}$ with
$\kappa=2.3(1)$.

At the critical healing rate $\mu_c^{cp}$ of the simple contact
process, the stationary density displays a weaker $\sigma$-dependence. A fit to a
power-law $\rho_{st} \sim \sigma^{\omega_{cp}}$ gives an exponent value of
$\omega_{cp}=0.108(2)$. In contrast, at the endpoint at healing rate $\mu^*$, the corresponding
exponent $\omega^* = 1.4(1)$ is larger than 1.

\begin{figure}
\includegraphics[width=\columnwidth]{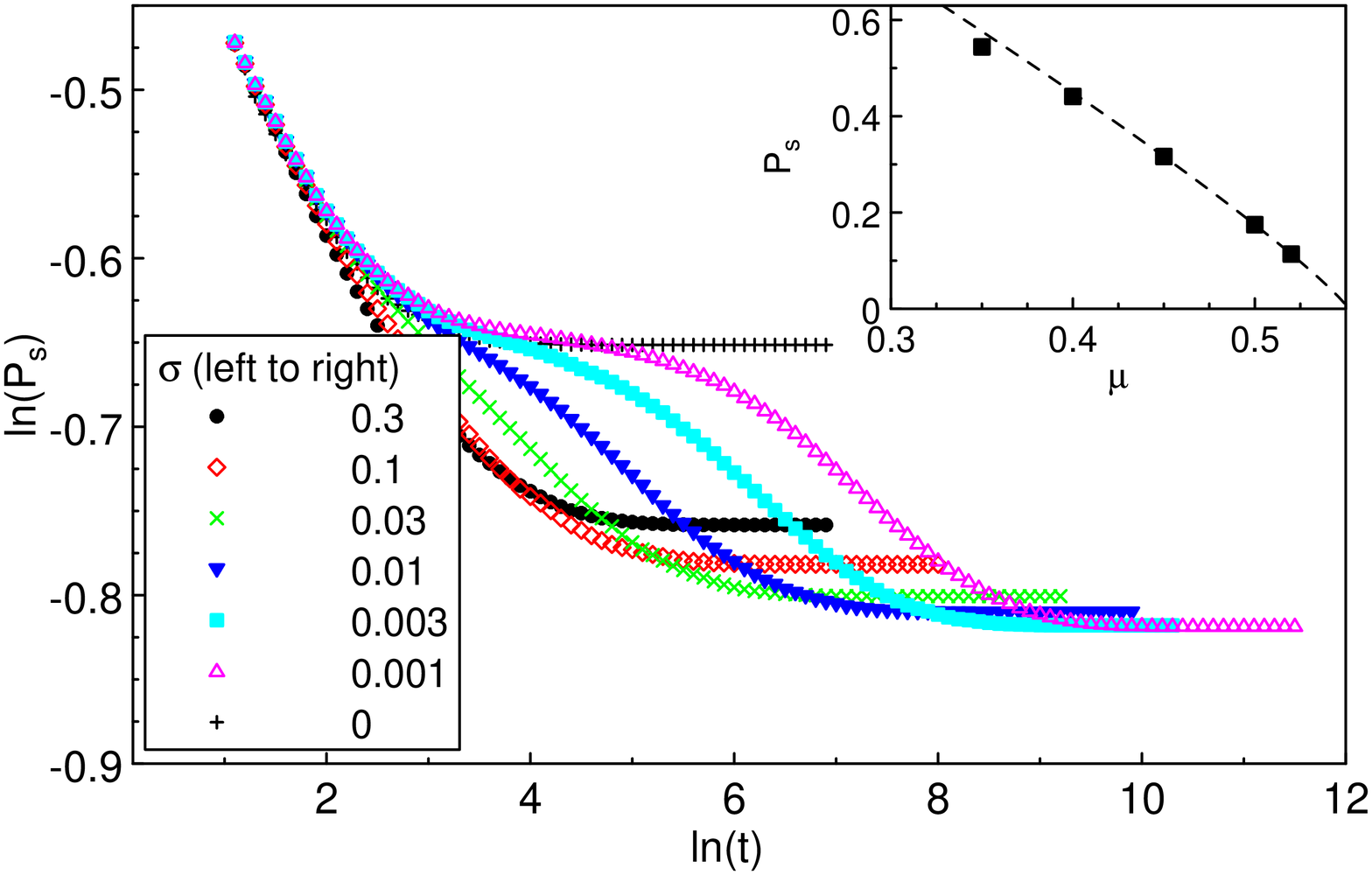}
\caption{(Color online) Spreading simulations: Survival probability $P_s$ as a function
of time $t$ at $\mu=0.4$ for various values of the boundary rate $\sigma$. The data are averages
over 100000 runs. Inset: Low-$\sigma$ limit of the stationary $P_s$ as a function of $\mu$. The dashed line
is a fit to $P_s \sim (\mu^* -\mu)^\beta$ with $\mu^*=0.552$ and $\beta=0.87(5)$ in agreement
with the PC universality class (see, e.g., Refs.\ \cite{Hinrichsen00,Odor04}).}
\label{fig:Ps_t}
\end{figure}

These results of the density decay simulations must be contrasted with those of the
spreading simulations. Figure \ref{fig:Ps_t} shows the time dependence of the survival
probability $P_s$ for $\mu=0.4$ and several $\sigma$.
At early times, all curves follow the $\sigma=0$ data due to the small values of
the rate of the boundary activation process (\ref{eq:rate_sigma}). (Note that the
$\sigma=0$ curve does \emph{not} reproduce the survival probability of the simple contact
process. This is because in our generalized contact process, a sample is
surviving as long as not every site is in state I$_1$ even if there are no active sites.)
In the long-time limit, the $P_s$ curves approach nonzero constants, as expected in an
active phase. However, in contrast to the stationary density $\rho_{st}$ (Fig.\ \ref{fig:rho_st}),
the stationary value of $P_s$ does not go to zero with vanishing boundary $\sigma$.
Instead, it approaches a $\sigma$-independent constant. We performed similar sets of
simulations at other values of $\mu$ in the range $\mu_c^{cp} < \mu < \mu^*$, with
analogous results. We therefore conclude that -- somewhat surprisingly --
the survival probability and the
stationary density of active sites display qualitatively different behavior at the
$\sigma=0$ phase transition.

We now show that the properties of these quantities can be understood within a simple
domain wall theory. The relevant long-time degrees of freedom at $\mu > \mu_c^{cp}$ and
$\sigma \ll 1$ are the domain walls between I$_1$ and I$_2$ domains. These domains are
formed during the early time evolution when the system follows the simple contact process
dynamical rules (\ref{eq:rate_barmu}) to (\ref{eq:rate_lambda}). At late times, the
domain walls can hop, they can branch (one wall branching into three), and they can
annihilate (two walls vanish if the meet on the same bond between two sites). This means,
the domain wall dynamics follows the branching-annihilating random walk with two
offspring (BARW2).

In our case, the BARW2 dynamics is controlled by two rates,
the domain wall hopping rate $\Gamma$ and the branching rate $\Omega$ (annihilation
occurs with certainty if two walls meet). These two rates depend on the underlying
generalized contact process dynamics. In the limit $\sigma \ll 1$ they are both linear in
the boundary rate, $\Gamma = \sigma F_\Gamma (\mu)$, $\Omega = \sigma F_\Omega(\mu)$
because a single boundary activation event is sufficient to start a domain wall hop or
branching ($F_\Gamma$ and $F_\Omega$ are nontrivial functions of $\mu$). Because both
rates are linear in $\sigma$, their ratio is $\sigma$-independent, thus the steady state
of the domain walls does not depend on $\sigma$ in the limit $\sigma \ll 1$. This
explains why the survival probability $P_s$ of the generalized contact process saturates
at a nonzero, $\sigma$-independent value in Fig.\ \ref{fig:Ps_t}. It also explains the
$\sigma$-dependence of the stationary density $\rho_{st}$ of active sites in
the following way: For $\sigma \ll 1$ and $\mu > \mu_c^{cp}$, active sites are created
mostly at the domain walls at rate $\sigma$. Consequently,
their stationary density is proportional to both $\sigma$ and the stationary domain
wall density $\rho_{dw}$, i.e., $\rho_{st} \sim \sigma \rho_{dw}$, in agreement with
Fig.\ \ref{fig:rho_st}. (The linear $\sigma$-dependence of $\rho_{st}$ is thus
\emph{not} due to the validity of mean-field theory.)

These results imply that the phase transition line at $\sigma=0$ between $\mu_c^{cp}$
and $\mu^*$ is \emph{not} a true critical line because there is no (nontrivial) diverging
length scale. It only appears critical because the stationary density of active sites
vanishes with $\sigma$. Note that this is also reflected in the fact that the system is not
behaving like a critical system right on the phase transition line $\sigma=0$
(no power-law time dependencies, for instance).
Instead, the physics of this transition line is controlled by the BARW2
dynamics of the domain walls with a finite correlation length for all $\mu_c^{cp} < \mu
< \mu^*$.

\subsection{Scaling at the contact process critical point $(\mu_c^{cp},0)$}
\label{subsec:CP_contact}

Even though the generalized contact process is not critical at $\sigma = 0$ and
$\mu>\mu_c^{cp}$, its behavior close to the critical point of the simple contact process
can be understood in terms of a phenomenological scaling theory.

Let us assume that the stationary density of active sites
close to $(\mu^{cp}_c,0)$ fulfills the homogeneity relation
\begin{equation}
\rho_{st}(\Delta \mu, \sigma) = b^{\beta_{cp}/\nu^{\perp}_{cp}} \rho_{st} (\Delta \mu\, b^{-1/\nu^{\perp}_{cp}}, \sigma b^{-y_{cp}})
\label{eq:rho_homogeneity}
\end{equation}
where $\Delta \mu = \mu - \mu_c^{cp}$ and $b$ denotes an arbitrary scale factor.
$\beta_{cp}$ and $\nu^\perp_{cp}$ are the usual order parameter and correlation
length exponents and $y_{cp}$ denotes the scale dimension of $\sigma$ at this
critical point. Setting $b=\sigma^{1/y_{cp}}$ then gives rise to the scaling form
\begin{equation}
\rho_{st}(\Delta \mu, \sigma) = \sigma^{\beta_{cp}/(\nu^\perp_{cp}y_{cp} )} X\left(
\Delta \mu\, \sigma^{-1/(\nu^\perp_{cp}y_{cp} )}\right)~
\label{eq:rho_scaling_cp}
\end{equation}
where $X$ is a scaling function.
At criticality, $\Delta\mu =0$, this leads to $\rho_{st}(0,\sigma) \sim \sigma^{\beta_{cp}/(\nu^\perp_{cp}y_{cp}
)}$ (using $X(0)=\textrm{const}$). Thus, $\omega_{cp}=
\beta_{cp}/(\nu^\perp_{cp}y_{cp})$. For $\sigma \to 0$ at nonzero $\Delta
\mu$, we need the large-argument limit of the scaling function $X$.
On the active side of the critical point, $\Delta \mu <0$, the scaling function must
behave as $X(x) \sim |x|^{\beta_{cp}}$ to reproduce the correct critical behavior
of the density, $\rho_{st} \sim |\mu-\mu_c^{cp}|^{\beta_{cp}}$.

More interesting is the behavior on the inactive side of the critical point, i.e., for
$\Delta \mu >0$ and $\sigma \to 0$. Here, we assume the scaling function to behave as
$X(x) \sim x^{-\kappa}$. In this limit, we thus obtain
$\rho_{st} \sim (\Delta \mu)^{-\kappa} \sigma^\omega$
(just as observed in Fig.\ \ref{fig:rho_st}) with $\omega=(\beta_{cp}+\kappa)/(\nu^\perp_{cp}y_{cp})$.
As a result of our scaling theory, the exponents $\omega, \omega_{cp}$ and $\kappa$ are not independent, they need to
fulfill the relation $\omega_{cp} (\beta_{cp} +\kappa) = \beta_{cp} \omega$.
Our numerical values, $\omega=1$, $\omega_{cp}=0.108(2)$ and $\kappa=2.32(10)$ fulfill this relation
in very good approximation, indicating that
they represent asymptotic exponents and validating the homogeneity relation
(\ref{eq:rho_homogeneity}). Using $\beta_{cp}=0.2765$ and $\nu^\perp_{cp}= 1.097$
\cite{Jensen99}, the resulting value for the scale dimension $y_{cp}$ of
$\sigma$ at the simple contact process critical point is $y_{cp}=2.34(4)$.

\subsection{The endpoint $(\mu^*,0)$}
\label{subsec:MCP}

Finally, we turn to the point $(\mu^*, \sigma)=(0.552,0)$ where the generic phase
transition line terminates on the $\mu$ axis. At first glance, one might suspect this
point to be a multicritical point because it is located at the intersection of two phase
transition lines. However, we argued in Sec.\ \ref{subsec:sigma0} (based on the domain wall theory)
that the
transition line at $\sigma=0$ and $\mu_c^{cp}<\mu< \mu^*$ is not critical. This implies
that the endpoint $(\mu^*,0)$ is not multicritical but a simple critical point in the
same universality class (\emph{viz.}, the PC class) as the generic transition at $\mu > \mu^*$.
In fact, the endpoint can be understood as the critical point of the BARW2 domain wall
dynamics in the limit $\sigma\to 0$.

To test this hypothesis, we first study the survival probability and density of active
sites as $\mu^*$ is approached along the $\mu$ axis. The inset of Fig.\ \ref{fig:Ps_t}
shows the stationary survival probability (more precisely, its saturation value for
$\sigma \to 0$) as a function of $\mu$. The data can be well fitted by a power-law
$P_s \sim (\mu^*-\mu)^\beta$ with $\beta=0.87(5)$.
The corresponding information on the stationary density of active
sites can be obtained from inset b) of Fig.\ \ref{fig:rho_st}. It shows the prefactor
$B_\mu$ of the linear $\sigma$-dependence $\rho_{st} = B_\mu \sigma$ as a function of
$\mu^* -\mu$. Sufficiently close to $\mu^*$, their relation can be fitted by a power law
$B_\mu \sim (\mu^*-\mu)^{\kappa^*}$ with $\kappa^*=0.91$.
Thus both $\beta$ and $\kappa^*$ agree with the order parameter exponent of the PC
universality class within their error bars. This confirms the validity of the domain wall
theory of Sec.\ \ref{subsec:sigma0} at $\mu^*$.

The discussion of the $\sigma$-dependence of $P_s$ and $\rho_{st}$ right at $\mu^*$
is somewhat more complicated because it is determined by the \emph{subleading}
$\sigma$-dependencies of the domain-wall rates $\Gamma$ and $\Omega$.
Moreover, because the dynamics is extremely slow at $\mu\approx\mu^*$ and
$\sigma \ll 1$, our numerical results close to the endpoint are less accurate
then our other results.
According to the domain wall theory of Sec.\ \ref{subsec:sigma0}, the stationary survival
probability should fulfill the homogeneity relation
\begin{equation}
P_s(\Delta \mu, \sigma ) = b^{\beta/\nu^\perp} P_s (\Delta \mu\, b^{-1/\nu^\perp}, \sigma
\, b^{-y^*})
\label{eq:Ps_homogeneity}
\end{equation}
where $\Delta \mu =\mu-\mu^*$ while $\beta$ and $\nu^\perp$ are the order parameter and
correlation length exponents of the BARW2 transition (PC universality class). The only
unknown exponent is $y^*$. The same homogeneity relation should hold for the domain
wall density, but \emph{not} the density of active sites.

Setting the scale factor to $b=\sigma^{1/y^*}$ gives
the scaling form
\begin{equation}
P_s(\Delta \mu, \sigma ) = \sigma^{\beta/(\nu^\perp y^*)} Y (\Delta \mu\,  \sigma^{-1/(\nu^\perp
y^*)})~.
\label{eq:Ps_scaling_mu*}
\end{equation}
Right at the endpoint, $\Delta \mu =0$, this gives $P_s \sim \sigma^{\beta/(\nu^\perp
y^*)}$. To test this power-law relation and to determine $y^*$, we performed spreading
simulations at $\mu=\mu^*$ and several $\sigma$ between 0.03 and 1. The low-$\sigma$
behavior (not shown) can indeed be fitted by a power law in $\sigma$ with an exponent
$\beta/(\nu^\perp y^*)=0.5(1)$. Using the well-known values $\beta=0.92$ and
$\nu^\perp=1.83$ of the PC universality class, we conclude $y^*=1.0(2)$. Within the
domain wall theory, $\rho_{DW} \sim P_s$ and the stationary density of active sites is
$\rho_{st} \sim \sigma \rho_{DW} \sim \sigma^{\omega^*}$ with $\omega^* =
1 + \beta/(\nu^\perp y^*)=1.5(1)$. This agrees well with the numerical estimate of 1.4(1)
obtained from the density decay simulations in Fig.\ \ref{fig:rho_st}.

The scaling form (\ref{eq:Ps_scaling_mu*}) can also be used to determine the shape of
the phase boundary at $\mu > \mu^*$. The phase boundary corresponds to a singularity of
the scaling function $Y$ at some nonzero value of its argument. Thus,
the phase boundary follows the power law $\sigma \sim (\mu-\mu^*)^{\nu^\perp y^*}$.
At fit of the data in Fig.\ \ref{fig:pd} leads to $\nu^\perp y^*=1.8(2)$ which implies
$y^*=1.0(1)$ in agreement with the above estimate from the spreading simulation data.

To investigate the time dependence of $P_s$ close to the endpoint, the homogeneity
relation (\ref{eq:Ps_homogeneity}) can be generalized to include a time argument. On the
right hand side, it appears in the scaling combination $(t/t_0) b^z$ with $t_0$ the
basic microscopic time scale. It is important to realize that this \emph{microscopic}
scale diverges as $\sigma^{-1}$ with $\sigma \to 0$ (independent of any criticality
at $\mu^*$). Thus, the right scaling combination is actually $t \sigma b^z$.
We used the resulting scaling theory to discuss the power-law decay of $P_s$ on the
phase boundary shown in Fig.\ \ref{fig:all_critical}a. The scaling theory predicts
$P_s \sim \sigma^{-\zeta} t^{-\delta}$ with $\zeta\equiv\delta$ as the endpoint is approached.
This agrees with our
numerical data (shown in the inset of Fig.\ \ref{fig:all_critical}a) which give
$\zeta \approx 0.284$

In summary, all our simulation data support the notion that the endpoint $(\mu^*,0)$
is a not a true multicritical point but a simple critical point in the same universality
class (PC) as the entire generic phase boundary at $\mu \ge \mu^*$. The behavior of some
observables makes it appear multicritical, though, because the microscopic time scale
of the domain wall dynamics diverges with $\sigma \to 0$.

\section{Conclusions}
\label{sec:conclusions}

In summary, we have studied the phase transitions of the generalized contact process
with two absorbing states in one space dimension by means of large-scale Monte-Carlo
simulations. We have found that this model has two different nonequilibrium phase
transitions, (i) the generic transition occurring for sufficiently hight values $\mu>\mu^*$
of the healing rate and nonzero values of the boundary activation rate $\sigma$, and
(ii) a transition at exactly $\sigma=0$ for $\mu_c^{cp}<\mu<\mu^*$.

The generic
transition is in the parity-conserving (PC) universality class (which coincides with the
DP2 class in one dimension) everywhere on the $\mu\ge\mu^*$ phase boundary, in agreement
with earlier work \cite{Hinrichsen97,HooyberghsCarlonVanderzande01}. In contrast, the
$\sigma=0$ transition turned out to be \emph{not} critical. The density of active sites
rather goes to zero with the vanishing boundary activation rate $\sigma$ while the survival
probability remains finite for $\sigma\to 0$. Its behavior is controlled by the BARW2
dynamics of the domain walls between different inactive domains (which is not critical
for $\mu_c^{cp}<\mu<\mu^*$). It is interesting to note that the behavior of our model at
$\sigma\equiv 0$ differs qualitatively from the $\sigma\to 0$ limit of the
finite-$\sigma$ behavior in the entire parameter region
$\mu_c^{cp}<\mu<\mu^*$.

As a result, the crossover between directed percolation (DP) critical behavior at
$\sigma\equiv 0$ and parity conserving (PC) critical behavior for $\sigma >0$ does not
take the naively expected simple scaling form. In particular, the generic ($\sigma>0$)
phase boundary does not continuously connect to the critical point of the $\sigma\equiv
0$ theory (the simple contact process critical point). Instead, it terminates at a separate
endpoint $(\mu^*,0)$ on the $\mu$-axis. While this point shares some characteristics with
a multicritical point, it is actually just a simple critical point in the same universality
class (PC) as the entire generic phase boundary.

We emphasize that the crossover between the DP and PC universality classes as a function
of $\sigma$ in our model is very different from that investigated by Odor and Menyhard
\cite{OdorMenyhard08}. These authors started from the PC universality class and
introduced perturbations that destroy the symmetry between the absorbing states or
destroy the parity conservation in branching and annihilating random walk models. They
found more conventional behavior that can be described in terms of crossover scaling. In
contrast, the transition rates (\ref{eq:rate_barmu}) to (\ref{eq:rate_sigma}) of our
model do not break the symmetry between the two inactive states anywhere in parameter
space.

Crossovers between various universality classes of absorbing state transitions have also
been investigated by Park and Park \cite{ParkPark07,ParkPark08,ParkPark09}.
They found a discontinuous jump in the phase boundary similar to ours along the so-called
excitatory route from infinitely many absorbing states to a single absorbing state
\cite{ParkPark07}. Moreover, there is some similarity between our mechanism and the so-called channel
route \cite{ParkPark08} from the PC universality class to the DP class which involves an
infinite number of absorbing states characterized by an auxiliary density. 
In our case, at $\sigma\equiv 0$ (but not at any finite $\sigma$), any configuration 
consisting of I$_1$ and I$_2$ only can be considered absorbing because active sites cannot be created. 
The density of I$_1$-I$_2$ domain walls then plays the role of the auxiliary density;
it vanishes at the endpoint $(\mu^*,0)$. However, our crossover
occurs in the opposite direction than that of Ref.\ \cite{ParkPark08}: The small
parameter $\sigma$ takes the system from the DP universality class to the PC class. Note
that an unexpected survival of active sites has also been observed in a version of the
nonequilibrium kinetic Ising model with strong disorder. Here, the disorder can
completely segment the system, and in odd-parity segments residual particles cannot decay
\cite{OdorMenyhard06}.

\begin{figure}[t]
\centerline{\includegraphics[width=8cm,clip]{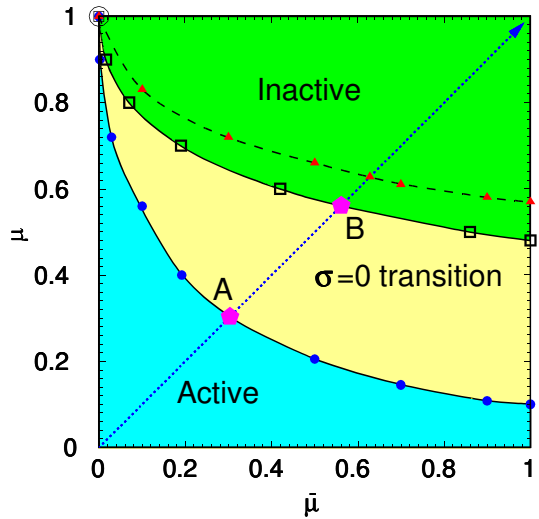}}
\caption{(Color online) Projection of the phase diagram of the generalized contact process
on the $\bar\mu-\mu$ plane. The individual symbols show the locations of the phase boundaries
as determined from our simulations: solid blue circles -- transition for $\sigma\equiv 0$
(simple contact process), solid red triangles -- generic transition for $\sigma=1$, open squares --
approximate location of the endpoint of the generic transition ($\sigma \to 0$) estimated
from the transition at $\sigma=0.01$. The lines are guides to the eye only. Points A and B are the
simple contact process critical point and the endpoint investigated in the main part of the paper.}
\label{fig:pd_barmu_mu}
\end{figure}

The generalized contact process as defined in eqs.\ (\ref{eq:rate_barmu}) to (\ref{eq:rate_sigma})
is characterized by \emph{three} independent rates (one rate can be set to one by rescaling the
time unit). In the bulk of our paper, we have focused on the case $\bar\mu = \mu$ for
which our system reduces to the usual contact process in the limit of $\sigma \to 0$.
In order to study how general our results are, we have performed a few simulation runs
for $\bar\mu \ne \mu$ focusing on the fate of the endpoint that separates the generic
transition from the $\sigma=0$ transition. The results of these runs are summarized in
Fig.\ \ref{fig:pd_barmu_mu} which shows the phase diagram projected on the $\bar\mu-\mu$
plane.
The figure shows that the line of endpoints of the generic phase boundary remains distinct from
the simple contact process ($\sigma=0$) critical line in the entire $\bar\mu-\mu$ plane. The
two lines only merge at the point $\bar\mu=0, \mu=1$ where the system behaves as compact
directed percolation \cite{Hinrichsen97}.

Our study was started because simulations at $\mu \gtrapprox \mu_c^{cp}$
and $\sigma \ll 1$ \cite{LeeVojta09} seemed to suggest that the generalized contact process with two
absorbing states is always active for any nonzero $\sigma$. The detailed work reported in this paper
shows that this is \emph{not} the case; a true inactive phase appears, but only
at significantly higher $\mu> \mu^*$. Motivated by this result, we also carefully reinvestigated
the generalized contact process with $n=3$ absorbing states which has been reported to be
always active (for any nonzero $\sigma$) in the literature
\cite{Hinrichsen97,HooyberghsCarlonVanderzande01}. However, in contrast to the
two-absorbing-states case, we could not find any inactive phase in this system.

Let us close by posing the question of whether a similar splitting between the $n=1$ critical
point and the $n=2$ phase transition line also occurs in \emph{other} microscopic models with
several absorbing states. Answering this questions remains a task for the future.

\section*{Acknowledgements}

We acknowledge helpful discussions with Geza Odor and Ronald Dickman. This work has been
supported in part by the NSF under grant no. DMR-0339147 and DMR-0906566 as well as by
Research Corporation.

\bibliographystyle{apsrev}
\bibliography{../00Bibtex/rareregions}
\end{document}